\documentclass[a4paper]{article}
\usepackage{amsmath}
\usepackage{amsfonts}
\usepackage{amssymb}
\usepackage{mathrsfs}
\usepackage{graphicx}
\usepackage{bm}
\usepackage{cite}
\usepackage{hyperref}

\input{tcilatex}
\begin{document}

\title{A model of quantum collapse induced by gravity}
\author{F. Lalo\"{e} \thanks{
laloe at lkb.ens.fr} \\
LKB, ENS-Universit\'{e} PSL, CNRS, 24 rue Lhomond, 75005\ Paris, France}
\date{\today }
\maketitle

\begin{abstract}
We discuss a model where a spontaneous quantum collapse is induced by the
gravitational interactions, treated classically. Its dynamics couples the
standard wave function of a system with the Bohmian positions of its
particles, which are considered as the only source of the gravitational
attraction. The collapse is obtained by adding a small imaginary component
to the gravitational coupling. It predicts extremely small perturbations of
microscopic systems, but very fast collapse of QSMDS (quantum superpositions
of macroscopically distinct quantum states) of a solid object, varying as
the fifth power of its size. The model does not require adding any
dimensional constant to those of standard physics.
\end{abstract}

\tableofcontents

\begin{center}
\ \ \ \ ********
\end{center}

A well-known difficulty in quantum mechanics is that the dynamical equations
(Schr\"{o}dinger or von Neumann equations) seem to predict the possible
occurrence of quantum superpositions of macroscopically distinct states
(QSMDS) \cite{Leggett-1980,Leggett-2002} that are never observed, for instance the creation of Schr\"{o}dinger cats \cite{Schrodinger-cat,Trimmer}. To solve the difficulty,
von Neumann \cite{von-Neumann} suggested to introduce a quantum collapse
postulate, which is nowadays part of most introductory textbooks on quantum
mechanics.

Several authors have proposed to relate quantum collapse to the effects of
gravity.\ One can for instance assume that gravity is the source of a random
noise acting on the state vector, and that this noise projects QSMDS onto
one of its components localized in space.\ In Refs. \cite{Diosi-1987,Diosi-1989}, Diosi discuss the introduction of
a stochastic term in the Schr\"{o}dinger (or von Neumann) equation that
efficiently destroys QSMDS; see also Refs. \cite{GGR-1990,Pearle-Squires}.\ In  Ref. \cite{Penrose-1996}, Penrose notes that the energy difference
associated with different mass distributions leads to a violation of energy
conservation, and suggest that this violation is spontaneously canceled by some random projection
mechanism. Other contributions may be found in Refs. \cite{Diosi-2014, Tilloy-Diosi, Adler-2016,Gasbarri-et-al-2017}. \ Reviews of this class of theories can be found
in \S ~III-B of Ref. \cite{Bassi-Lochan-Satin-Singh-Ulbricht} and in Ref. \cite{Singh-2015}.

Here we propose a model of the quantum dynamics that also provides a collapse, but with equations that are completely deterministic; gravity is treated as a
classical field originating from the Bohmian positions of the particles. In
classical physics, gravity already plays special role, since it determines
the curvature of space-time.\ In our model, we attribute to gravity another
special feature, which is to introduce small non-Hermitian component in the
evolution equation of the state vector.\ Nothing in this model is
stochastic; the only source of randomness is the initial randomness of the
Bohmian position, as in the de Broglie-Bohm (dBB) theory \cite%
{de-Broglie,Bohm,Holland, Bacchiagaluppi-Valentini,Duerr-et-al,Bricmont}.
This model is in the line of a general view where space-time remains
classical, and where the source of the curvature of space-time is the
Bohmian positions of the particles; the various quantum fields propagate inside this
classical space-time frame.

Combining elements from dBB and spontaneous collapse \cite{GRW,CSL} theories is not a completely new idea. Ref. \cite{Allori-2008} proposes to localize the wave function around the Bohmian positions, but with no real change of the Schr\"{o}dinger dynamics;  moreover, gravity plays no role in the localization process. Refs. \cite{Bedingham-2011} and \cite{Tumulka-2011} consider a back action of the Bohmian positions on the wave function, but with a stochastic term, as in standard spontanous collapse theories \cite{GRW,CSL}.

For the sake of simplicity, here we discuss only spinless non-relativistic
particles (including spins within a Pauli theory is nevertheless not
particularly difficult). As in Refs. \cite{SDAP,ABD}, we use a dynamics
involving an \textquotedblleft expanded description\textquotedblright\ of
the physical system: to the standard wave function $\Psi $ defined in the
configuration space, we add (in the same space) a mathematical point $Q$,
whose coordinates are determined by the Bohmian positions $q_{n}$ of all its
particles.\ Incidentally, and in contrast with the usual interpretations of
the de Broglie-Bohm (dBB) theory, we make no particular assumption
concerning the physical reality of these positions; they can be seen, either
as physically real, or as a pure mathematical object appearing in the
dynamical equations.

The equations of this dynamics are given in \S ~\ref%
{modified-quantum-dynamics}. In \S ~\ref{discussion} we discuss the
predictions of the model in various situations, showing that it introduces
no significant change for microscopic systems while it rapidly projects
QSMDS onto one of its localized components. A conclusion is given in \S ~\ref%
{conclusion}.

\section{Modified quantum dynamics}

\label{modified-quantum-dynamics}

We assume that the Hamiltonian $H$ of a physical system is the sum of its
internal Hamiltonian $H_{\text{int}}$ (including the kinetic energy of the
particles and their mutual interactions) and of a gravitational Hamiltonian $%
H_{G}$, due to the attraction of external masses with mass density $n_{G}(%
\mathbf{r})$:
\begin{equation}
H=H_{\text{int}}+H_{G}  \label{grav-1}
\end{equation}%
with:%
\begin{equation}
H_{G}=-gGm \int \text{d}^{3}r~\Psi ^{\dagger }(\mathbf{r})\Psi (\mathbf{r}%
)\int \text{d}^{3}r^{\prime }~\frac{1}{\left\vert \mathbf{r}-\mathbf{r}%
^{\prime }\right\vert }n_{G}(\mathbf{r}^{\prime })  \label{grav-2}
\end{equation}%
In this relation, $g=1$ in standard theory, $G$ is Newton's constant, $m$
the mass of the particles, and $\Psi (\mathbf{r})$ the quantum field
operator of the particles contained in the physical system. With external
sources of gravity, this Hamiltonian is completely standard. If $n_{G}(\mathbf{r}^{\prime }) $ is set equal to quantum local density average $< \Psi ^{\dagger }(\mathbf{r})\Psi (\mathbf{r}%
) >$, we obtain the usual Schr\"{o}dinger-Newton equation \cite{Bahrami-Bassi-et-al}.

\subsection{Evolution of the state vector}

We now leave standard quantum mechanics by making two non-standard
assumptions.\ First, we assume that $H_{G}$ actually describes the internal
gravitational attraction of the system, and that $n_{G}(\mathbf{r})$ is
determined by the Bohmian positions $q_{n}$ of the $N$ particles of the system:%
\begin{equation}
n_{G}(\mathbf{r})=m\sum_{n=1}^{N}\delta (\mathbf{r}-\mathbf{q}_{n})  \label{grav-3}
\end{equation}%
Incidentally, one could also perform a spatial average over a distance $a_{L}$, as usual
in GRW and CSL \cite{GRW,CSL} theories, and write for instance:%
\begin{equation}
n_{G}(\mathbf{r})=\frac{m}{\pi ^{3/2}a_{L}^{3}}\sum_{n=1}^{N}~\text{e}^{-(\mathbf{r%
}-\mathbf{q}_{n})^{2}/\alpha _{L}^{2}}  \label{grav-4}
\end{equation}%
Nevertheless, in what follows, we will only use the simpler form (\ref%
{grav-3}). Similarly it has been proposed in Ref. \cite{Prezdho-Brooksby} to study chemical reactions (within standard quantum mechanics) by an approximation where the nuclei are treated classically, and where the backreaction of the quantum electrons on the nuclei is obtained by sampling the Bohmian positions of the electrons over their quantum distribution.

Second, as in Ref. \cite{Diosi-Papp-2009}, we assume that the dimensionless constant $g$ has a small
imaginary part $\varepsilon $:%
\begin{equation}
g=1-i\varepsilon   \label{grav-5}
\end{equation}%
(one could choose any small number, for instance $\varepsilon =\alpha $, the
fine structure constant).\ This introduces an antiHermitian part in $H_{G}$:%
\begin{equation}
H_{G}=H_{G}^{0}+iL  \label{grav-5-bis}
\end{equation}%
where $H_{G}^{0}$ is the Hermitian part of $H_{G}$:%
\begin{equation}
H_{G}^{0}=H_{G}(\varepsilon =0)  \label{grav-5-ter}
\end{equation}%
and where $L$ is the localization operator:%
\begin{equation}
L=\varepsilon Gm\int \text{d}^{3}r~\Psi ^{\dagger }(\mathbf{r})\Psi (\mathbf{%
r})\int \text{d}^{3}r^{\prime }~\frac{1}{\left\vert \mathbf{r}-\mathbf{r}%
^{\prime }\right\vert }n_{G}(\mathbf{r}^{\prime })  \label{grav-6}
\end{equation}%
This operator is diagonal in the position representation. It is  the second quantized form of the sum of $N$ single particle potentials taking large values in the vicinity of the Bohmian positions, in the regions of space where the gravitational attraction by these positions is strong.
With (\ref{grav-3}), this expression becomes:%
\begin{equation}
L=\varepsilon Gm^{2}\int \text{d}^{3}r~\Psi ^{\dagger }(\mathbf{r})\Psi (%
\mathbf{r})~\sum_{n=1}^{N}\frac{1}{\left\vert \mathbf{r}-\mathbf{q}_{n}\right\vert
}  \label{grav-6-bis}
\end{equation}%
In \cite{SDAP} we introduced a localization operator where the quantum operator
$\Psi ^{\dagger }(\mathbf{r})\Psi (\mathbf{r})$ is coupled to the Bohmian
positions with a Gaussian spatial average of range $a_{L}$. Here the
Gaussian spreading function is replaced by a gravitational type of coupling that is
proportional to the inverse distance.

The state vector $\left\vert \Phi (t)\right\rangle $ evolves according to:%
\begin{equation}
i\hbar \frac{\text{d}}{\text{d}t}\left\vert \Phi (t)\right\rangle =\left[ H_{%
\text{int}}+H_{G}\right] \left\vert \Phi (t)\right\rangle  \label{grav-7}
\end{equation}%
If $\varepsilon \neq 0$, the norm of $\left\vert \Phi (t)\right\rangle $
does not remain constant.\ We can nevertheless introduce the normalized ket $%
\left\vert \overline{\Phi }(t)\right\rangle $:%
\begin{equation}
\left\vert \overline{\Phi }(t)\right\rangle =\frac{1}{\sqrt{\left\langle
\Phi (t)\right. \left\vert \Phi (t)\right\rangle }}\left\vert \Phi
(t)\right\rangle  \label{grav-10}
\end{equation}%
and set:%
\begin{equation}
D_{\Phi}(\mathbf{r})=\left\langle \overline{\Phi }(t)\right\vert \Psi
^{\dagger }(\mathbf{r})\Psi (\mathbf{r})\left\vert \overline{\Phi }%
(t)\right\rangle  \label{grav-9}
\end{equation}%
This normalized state then evolves according to:%
\begin{equation}
i\hbar \frac{\text{d}}{\text{d}t}\left\vert \overline{\Phi }(t)\right\rangle
=\left[ H_{\text{int}}+H_{G}^{0} + i\varepsilon Gm \int \text{d}^{3}r\int
\text{d}^{3}r^{\prime }~\left[ \Psi ^{\dagger }(\mathbf{r})\Psi (\mathbf{r}%
)-D_{\Phi}(\mathbf{r})\right] ~\frac{1}{\left\vert \mathbf{r}-\mathbf{r}%
^{\prime }\right\vert }\;n_{G}(\mathbf{r}^{\prime })\right] \left\vert \overline{\Phi }(t)\right\rangle  \label{grav-11}
\end{equation}

To summarize, the two non-standard ingredients of our model are:

- the use of the Bohmian positions to define a density of matter in ordinary
space; this density is the source of the classical gravitational field
involving the usual Newton constant $G$.

- the introduction of a small imaginary part in $G$, so that the dynamics
becomes irreversible and collapses QSMDS, as we see below.

\subsection{Evolution of the Bohmian positions}

We assume that the Bohmian positions $\mathbf{q}_{n}$ evolve according to
the usual Bohmian equation of motion:%
\begin{equation}
\frac{d\mathbf{q}_{n}\left( t\right) }{dt}=\frac{\hslash }{m} ~ \overrightarrow{%
\bigtriangledown }_{n}\xi \left( \mathbf{r}_{1},\mathbf{r}_{2},..,\mathbf{r}%
_{N}\right)   \label{sdap-5-ter}
\end{equation}%
where $\xi \left( \mathbf{r}_{1},\mathbf{r}_{2},..,\mathbf{r}_{N}\right) $
is the phase of the wave function $\Phi \left( \mathbf{r}_{1},\mathbf{r}%
_{2},..,\mathbf{r}_{N}\right) $, and $\overrightarrow{\bigtriangledown }_{n}$
the gradient taken with respect to $\mathbf{r}_{n}=\mathbf{q}_{n}$. Equivalently, this
equation can also be written:%
\begin{equation}
\frac{d\mathbf{q}_{n}\left( t\right) }{dt}=\frac{\hslash }{im~D_{\Phi }(\mathbf{r})}%
\left\langle \overline{\Phi }(t)\right\vert \Psi ^{\dagger }(\mathbf{r})%
\bm{\nabla}_{\mathbf{r}}\Psi (\mathbf{r})- \bm{\nabla}_{\mathbf{r}%
}\Psi^{\dagger} (\mathbf{r}) \Psi (\mathbf{r})\left\vert \overline{\Phi }%
(t)\right\rangle  \label{sdap-5-quater}
\end{equation}%
The condition of \textquotedblleft quantum
equilibrium\textquotedblright\  means that, when averaged over many
realizations of an experiment, the distribution of the Bohmian positions  in configuration space
coincides with the modulus square of the wave function.\ In standard dBB theory with the usual Schr\"{o}dinger equation, if this condition is satisfied at the initial time, it is also satisfied at any time. But this property no longer holds in our case, since we have
modified the dynamics of the wave function. Nevertheless, in Ref. \cite{ABD}
we discuss why the relaxation process studied by Towler, Russell and
Valentini \cite{Valentini-2005, Valentini-2012} should ensure that this
condition is still valid to an excellent approximation, except in a very
short transient time during the appearance (and almost immediate collapse)
of a QSMDS.

\section{Discussion}

\label{discussion}

We now discuss the effect of the localization term on the state vector.\ The
situation is similar to that already considered in Refs. \cite{SDAP,ABD}
except that, here, the time constants of the collapse mechanism arise from a
gravitational energy coupling the quantum particles with their Bohmian
positions. We discuss only the simpler version (\ref{grav-3}) of the model,
which introduces no fundamental parameter $a_{L}$, but similar conclusions
apply as well if a non-zero value of $a_{L}$ is chosen.

\subsection{Negligible effects on microscopic systems}

\label{microscopic}

Consider first the non-relativistic Schr\"{o}dinger equation of the electron
and proton in a Hydrogen atom, ignoring the spins for the sake of simplicity.\ Each of
the two particles is subjected to two attractions:

-- the usual Coulomb attraction, which introduces the usual two-body
potential in the Schr\"{o}dinger equation for the wave function.

-- the gravitational attraction, appearing as a one-body attractive
potential towards the position of an additional variable: the electron is
attracted towards the Bohmian position $\mathbf{q}_{p}$ of the proton, and
conversely the proton is attracted towards the Bohmian position $\mathbf{q}%
_{e}$ of the electron.

The ratio $X$ between the Coulomb and gravitational interactions is very
large:%
\begin{equation}
X\simeq \frac{q^{2}}{4\pi \varepsilon _{0}}\frac{1}{Gm_{e}m_{p}}\simeq
10^{39}  \label{grav-15}
\end{equation}%
where $\varepsilon _{0}$ is the permittivity of vacuum, $q$ the electronic
charge, $m_{e}$ the mass of the electron and $m_{p}$ the mass of the proton.
This enormous value of $X$ ensures that the gravitational component plays no
role in practice: we just recover the well-known fact that the gravitational
attraction remains completely negligible in the Hydrogen atom.\ The
divergences of $n_{G}(\mathbf{r})$ when $\mathbf{r}=\mathbf{q}_{n}$ do not
create any special problem: as in the standard theory of the Hydrogen atom,
they only introduce kinks in the wave function, but these kinks are $10^{29}$
times less pronounced than those introduced by the Coulomb potential; in
practice, they have no effect.\ Moreover, the statistical distribution of $%
\mathbf{q}_{p}$ and $\mathbf{q}_{e}$ over many realizations coincides with
the corresponding quantum distributions.\ Clearly, changing in this way the
center of gravitational attraction has no practical consequence.\ In
addition to this change, the model introduces a small imaginary component to
the gravitational part of the Hamiltonian, which introduces an even more
negligible perturbation.

Another example illustrates why, in most cases, the localization term has a
very small effect.\ If $\left\vert \overline{\Phi }(t)\right\rangle $ is an
eigenstate of the Hamiltonian $H_{\text{int}}+H_{G}^{0}$, the average energy
$\left\langle H_{\text{int}}+H_{G}^{0}\right\rangle $ remains constant:
\begin{equation}
\frac{\text{d}}{\text{d}t}\left\langle H_{\text{int}}+H_{G}^{0}\right\rangle
=0  \label{grav-16}
\end{equation}%
More generally, if $\left\vert \overline{\Phi }(t)\right\rangle $ is an
eigenstate of $A$ at time $t$, the localization term has no effect on the
derivative of the average value of $A$ at time $t$:%
\begin{equation}
\left. \frac{\text{d}}{\text{d}t}\right\vert _{\text{loc}}\left\langle
\overline{\Phi }(t)\right\vert A\left\vert \overline{\Phi }(t)\right\rangle
=0  \label{grav-17}
\end{equation}%
This is because, if $a$ is the eigenvalue of $A$, we have:%
\begin{align}
\left. \frac{\text{d}}{\text{d}t}\right\vert _{\text{loc}}\left\langle
\overline{\Phi }(t)\right\vert A\left\vert \overline{\Phi }(t)\right\rangle
& =\frac{2\varepsilon Gm}{\hslash }\int \text{d}^{3}r\int \text{d}%
^{3}r^{\prime } ~\left\langle \overline{\Phi }(t)\right\vert \left[ \Psi
^{\dagger }(\mathbf{r})\Psi (\mathbf{r})-D_{\Phi}(\mathbf{r})\right]
A\left\vert \overline{\Phi }(t)\right\rangle ~\frac{1}{\left\vert \mathbf{r}-%
\mathbf{r}^{\prime }\right\vert }\;n_{G}(\mathbf{r}^{\prime })  \notag \\
& =\frac{2\varepsilon aGm}{\hslash }\int \text{d}^{3}r\int \text{d}%
^{3}r^{\prime }~\left[\left\langle \overline{\Phi }(t)\right\vert \Psi
^{\dagger }(\mathbf{r})\Psi (\mathbf{r})\left\vert \overline{\Phi }%
(t)\right\rangle -D_{\Phi}(\mathbf{r})\right] ~\frac{1}{\left\vert \mathbf{r}%
-\mathbf{r}^{\prime }\right\vert }\;n_{G}(\mathbf{r}^{\prime })=0
\label{grav-18}
\end{align}%
The situation is therefore different from that obtained with GRW and CSL
theories \cite{GRW,CSL}, where the localization mechanism constantly
transfers energy to all particles at a small rate: in our model, if the
system is in a stationary state, thermal equilibrium for instance, its
energy remains constant. The reason for this difference is that, in GRW and
CSL theories, the random localization process involves a noise that is
discontinuous in time, and therefore has a very broad spectrum (infinite in the case of a
Wiener process); it cannot be treated as a first order perturbation and, for instance, the
Ito term has to be included.\ In our model, the localization term is
continuous and has a limited frequency spectrum (determined by the motion of
the Bohmian positions); since the coupling constant is very small, it can be
treated by first order perturbation theory, and has a much softer effect.

\subsection{Fast resolution of QSMDS}

\label{resolution}

Assume now that the quantum state describes a QSMDS situation, for instance
a measurement pointer (or any macroscopic object) in a superposition of two quantum states localized in two different regions of space. By contrast, the Bohmian positions remain grouped
together, forming a cluster that occupies only one of these regions of space. Therefore, in the two branches
of the state vector, a strong mismatch then occurs between the quantum density of
particles and the Bohmian density (but with a different sign), so that the effect of the localization
operator $L$ on these branches is significantly different.\ To evaluate its consequences we can, in (\ref{grav-11}),
ignore the normalization term in $D_{\Phi}(\mathbf{r})$, which
affects both branches in the same way and does not change their relative
amplitude. In the \textquotedblleft full component\textquotedblright\ where
the Bohmian density accompanies the quantum density, the localisation term in the right hand side of (\ref{grav-7}) multiplies the wave
function by a number that is of the order of (half of the absolute value of) the self-gravitational
energy $E_{\text{sg}}$ of the pointer, multiplied by the constant $%
\varepsilon $; in the \textquotedblleft empty component\textquotedblright ,
it multiplies the wave function by an energy that is negligible with respect
to this self-gravitational energy.\ Altogether, the differential effect
takes place with a time constant of the order of:%
\begin{equation}
\tau _{\text{collapse}}\simeq \frac{\hslash }{\varepsilon |E_{\text{sg}}|}
\label{grav-12}
\end{equation}%
with:%
\begin{equation}
|E_{\text{sg}}|\simeq G\frac{M^{2}}{L}  \label{grav-13}
\end{equation}%
where $M$ is the mass of the pointer and $L$ its size (we assume that the
two wave packets of the pointer are separated by approximately its size, or more).
If, for instance, $L=0.1$ mm and $M=10^{-6}$ g, and assuming $\varepsilon
=10^{-3}$, we find:%
\begin{equation}
\tau _{\text{collapse}}\simeq 10^{-6}\text{ s}  \label{grav-14}
\end{equation}

We note that $E_{\text{sg}}$ varies as the fifth power of the size of the
pointer (at constant density). For instance, if $L=1~\mu$m, we obtain a long
collapse time\ $\tau _{\text{collapse}}\simeq 10^{4}$ s.\ In experiments
such as those of Ref.\ \cite{Arndt-et-coll}, very large molecules could fly
on different path without being collapsed if the duration of the flight is
shorter than this time.\ The model thus
predicts a relatively sharp border between small objects that can reach and stay in a quantum superposition of remote states, and larger ones that almost immediately get projected onto one single
location.\ As discussed in \cite{ABD}, the origin of this projection is the
cohesive internal force of solid objects, which forces the Bohmian positions to remain clustered together; gases that do not have this
internal cohesion do not undergo the same effect. Interestingly, in the
correlated worldline (CWL) theory of quantum gravity \cite%
{Stamp-2015,Stamp-2016}, fifth powers of the masses also appear in the
mutual binding energy for paths.


\subsection{Large systems}
\label{large}

In most situations (except, of course, during the appearance of a QSMDS),
the space distribution of Bohmian variables accurately coincides with the
quantum space distribution $D_{\Phi }(\mathbf{r})$.\ Assuming that the
gravitational attraction originates from the distribution of Bohmian
positions is not very different than assuming that the source of attraction
is the quantum distribution $D_{\Phi }(\mathbf{r})$. The effect of the
localization term will then just be to (slowly) localize the macroscopic
system inside itself, or to move towards region of lower gravitational potential. This term should have no observable effect, except
maybe on very long time and space scales such as those considered in astrophysics; its effect is somewhat reminiscent of the attraction of the so called dark matter.

We note in passing that macroscopic quantum superpositions of states that do not produce  different spatial distributions of masses are not reduced by the localization process of the model. For instance, if the flow of electrons in a superconducting ring is in a superposition of two states having rotations in opposite directions, no significant collapse takes place. Fast collapse occurs only to resolve QSMDS involving different gravitational fields, as suggested by Penrose \cite{Penrose-1996}.

%

\subsection{Measurements}

When an apparatus $M$ is used to measure a quantum system $S$, both physical systems become entangled under the effect of their mutual interaction. The state vector then splits into several branches, each containing a state of $S$ that is an eigenstate of the measured observable. During the first stages of measurement, as long as the entanglement remains microscopic, the localization term plays no special role. But, when the entanglement involves states of $M$ involving significantly different distributions of masses in space, for instance different positions of a pointer, then the fast collective collapse takes place: all branches but one of the state vector vanish. The collapse process is therefore initiated inside the measurement apparatus, but immediately propagates back to $S$ by a standard quantum nonlocal effect. This is, for instance, what happens in a Bell experiment. No collapse therefore occurs before a significant part of the measurement apparatus  $M$ is part of the entanglement. The result of measurement is determined by the initial Bohmian positions of all particles and, as discussed in \cite{Tastevin-Laloe}, in some cases the result is primarily determined by the initial Bohmian positions of the measurement apparatus.

This scenario fits rather well with an old quotation by Pascual Jordan \cite{Jordan}): \textquotedblleft observations not
only disturb what has to be measured, they \textit{produce} it.\ In a
measurement of position, the electron is forced to a decision.\ We compel it
to assume a definite position; previously it was neither here nor there, it
had not yet made its decision for a definite position...\textquotedblright.

\subsection{No signaling}

When introducing nonlinearities in quantum dynamics, one should be careful about avoiding superluminal communications \cite{Gisin,Bassi-Hejazi,Diosi-2016}. In the GRW
\cite{GRW} and CSL \cite{CSL} versions of modified Schr\"{o}dinger dynamics,
nonlinearity and stochasticity compensate each other to cancel superluminal
signaling. Similarly, the nonlinear Schr\"{o}dinger-Newton equation can be made compatible with the no-signaling requirement by changing it to a stochastic differential equation \cite{Nimmrichter-2015}. Here, the situation is somewhat different: the nonlinearity is
not introduced as a term coupling the state vector directly to itself, but
by the reaction of Bohmian positions onto the wave function; the
stochasticity does not arise  from a
random process constantly acting on the wave function, but from the random values of the initial Bohmian positions.

In order to ensure that  the
Hamiltonian $H_{G}(\varepsilon =0)$ is nonsignaling, we  introduce a retarded potential into in Eq. (\ref%
{grav-2}):%
\begin{equation}
n_{G}(\mathbf{r}^{\prime })\Rightarrow n_{G}(\mathbf{r}^{\prime },t-\frac{%
\left\vert \mathbf{r}-\mathbf{r}^{\prime }\right\vert }{c})  \label{grav-19}
\end{equation}%
where $c$ is the speed of light. We then just have to
check that the localization term proportional to $\varepsilon $ is also nonsignaling.

Assume that the system, described by the density operator $\rho
(t)=\left\vert \Psi (t)\right\rangle \left\langle \Psi (t)\right\vert $, is
made of two remote subsystems $A$ and $B$, respectively occupying regions of
space $S_{A}$ and $S_{B}$, and described by the partial density operators $%
\rho _{A}(t)$ and $\rho _{B}(t)$. We denote $\left\{ \left\vert
n_{A}\right\rangle \right\} $ an ensemble of states of $A$  providing an
orthonormal basis, and $\left\{ \left\vert n_{B}\right\rangle \right\} $ a
similar basis for system $B$; for instance, $n_{A}$ and $n_{B}$ are abbreviated
notations for the positions of the $N_{A}$ particles that are inside $S_{A}$, and $N_B$ particles
 inside $S_{B}$, repectively.\ The evolution of the matrix elements of $%
\rho _{A}(t)$\ introduced by the localization term in $\varepsilon $ is
given by:%
\begin{align}
\left. \frac{\text{d}}{\text{d}t}\right\vert _{\text{loc}}& \left\langle
n_{A}\right\vert \rho _{A}(t)\left\vert n_{A}^{\prime }\right\rangle  \notag
\\
& =\frac{2\varepsilon Gm}{\hslash }\sum_{n_{B}}\left\langle
n_{A},n_{B}\right\vert \int \text{d}^{3}r\int \text{d}^{3}r^{\prime }~\left[
\Psi ^{\dagger }(\mathbf{r})\Psi (\mathbf{r})-D_{\Phi }(\mathbf{r}),\rho (t)%
\right] _{+}~n_{G}(\mathbf{r}^{\prime },t-\frac{\left\vert \mathbf{r}-%
\mathbf{r}^{\prime }\right\vert }{c})\left\vert n_{A}^{\prime
},n_{B}\right\rangle  \label{grav-20}
\end{align}%
where $\left[ C,D\right] _{+}$ denotes the anticommutator of $C$ and $D$.

We now assume that system $A$ is microscopic, but that $B$ is macroscopic, and that at some time it is driven to a QSMDS, for instance because a quantum measurement is
performed in this region $B$.\ We are interested in the possible
effects  on the
partial density operator $\rho _{A}(t)$ of the resolution of this QSMDS by the localization operator. The operator in the
right hand side of (\ref{grav-20}) contains the sum of four contributions: $%
L_{AA}$, $L_{BB}$, $L_{AB}$ and $L_{BA}$.\ Here the first index $A$ (or $B$)
indicates that the integration variable $\mathbf{r}^{\prime }$ lies in region $%
S_{A}$ (or $S_{B}$), which determines the source of localization; the
second index indicates that the integration variable $\mathbf{r}$ lies in
region $S_{A}$ (or $S_{B}$), which determines the target of the localization
process.\ Since we assume that $A$ is microscopic, we can ignore $L_{AA}$, which
is local and remains extremely small since $A$ is microscopic.\ We are
actually only interested in the terms having a macroscopic source, in other words in the effects of $L_{BB}$ and $L_{BA}$.

In fact, $L_{BB}$ is clearly the most important term. It looks local
since it corresponds to a localization occurring entirely in region $S_{B}$ by the collective spontaneous localization process discussed in \S~\ref{resolution};
nevertheless, since it acts on the density operator $\rho (t)$, which is a
nonlocal object if systems $A$ and $B$ are in an entangled state, this term can
introduce quantum nonlocality (in particular violations of the Bell inequalities). For
instance, if the measurement is performed on two spin $1/2$ particles in a
singlet spin state, as soon as a single spin measurement is performed in region $S_{B}$
along a direction $\mathbf{u}$, the
localization term will cancel one component of the singlet state; which component is cancelled depends on the result of measurement.\ In other words, in a single realization of the experiment, the spin
state in region $S-A$ will immediately be projected onto the opposite spin state on the same
direction $\mathbf{u}$. It is nevertheless well-known that this nonlocatity
does not imply any possible superluminal communication -- this is the famous
\textquotedblleft peaceful coexistence between quantum mechanics and
relativity\textquotedblright\ \cite{Shimony-peaceful-coexistence}.\ Indeed, if we
consider the average over many realizations, the density matrix of system $A$
remains completely independent of $\mathbf{u}$.\ Technically, while in the right hand side of (%
\ref{grav-20}) $n_{G}$ fluctuates in region $B$ from one realization to the
next, on average it can be replaced by the local density associated with the standard (non-collapsed) solution of the Schr\"{o}dinger equation; this provides the average effect of the localization  on the density operator of $A$. So, term $%
L_{BB}$ ensures that we recover  the usual nonlocal quantum correlations
between the remote subsystems $A$\ and $B$, the violation of the Bell
inequalities, etc., but without any superluminal communication.

We finally have to consider the effect of the term $L_{BA}$. It also
implies that the measurement result obtained in region $S_{B}$ may influence
the evolution of the density operator $\rho _{A}(t)$, but the
effect is much weaker that that of $L_{BB}$ since it tends to zero when the distance between regions $%
S_{A}$ and $S_{B}$ increases.\ Again, the average of this effect
over many realizations is obtained by replacing $n_{G}$ by the standard quantum density in space. It
is not signaling because of the delay $\left\vert \mathbf{r}-\mathbf{r}%
^{\prime }\right\vert /c$ appearing in the right hand side of (\ref{grav-20}): whatever
is done to change the Bohmian density inside subsystem $B$ cannot affect the
evolution of subsystem $A$ at any time earlier than the minimum delay required by
relativity.

For the sake of simplicity, we have assumed that $A$ is mùicroscopic and $B$ macroscopic, but the discussion could easily be generalized to the case where both are macroscopic. Our general conclusion, therefore, is that the model is nonsignaling, at least in all situations that we have considered.

\subsection{Differences and similarities with GRW/CSL theories}
\label{differences}

In the dynamical equations of the model, we have assumed that the Bohmian position of every particle is the source of gravity acting on all other particles. This is of course necessary for the Hermitian part of the Hamiltonian (obtained with $\varepsilon = 0$) if one wishes to reproduce the usual effects of gravity. But we have also assumed that this is  true for the antihermitian term (term in $\varepsilon$), introducing in this way \textquotedblleft mutual collapse terms\textquotedblright . As a consequence, our localization term in the dynamical equation is similar to a two-body  interaction term. By contrast, the localization term of GRW or CSL theories is rather described by a single-particle potential: the state vector is subjected to the effect of random localization terms acting on all
particles independently, with a probability rule  that depends on the values of
the wave function at the positions of all particles. In other words, in our model the collapse is a collective effect, by contrast with GRW/CSL theories. This difference has several consequences.

A first consequence is that, within our model, the localization rate
varies roughly proportionally to the square of the number of particles involved in a
QSMDS.\ Therefore, much smaller values of the collapse coupling constant can
be used, without losing a very fast collapse rate of QSMDS.\ In particular, this explains
why the undesirable heating effects initially predicted in~\cite{Diosi-1989}
with a gravitational collapse do not occur here.

Another consequence is
that, as discussed in \S ~\ref{microscopic}, the localization process of
this model is intrinsically softer than that of GRW and CSL, which have a
very short correlation time and therefore a broad spectrum (actually
infinitely broad); here, the gravitational attraction towards Bohmian position is
continuous in time, so that it can be treated perturbatively to first order
(for instance, it does not introduce Ito terms). As discussed in \S~\ref{large}, for large solid bodies, we obtain an effect of localization that is much weaker than that of GRW/CSL theories \cite{ABD}, so that it should be more difficult to detect experimentally (each particle in a solid is localized only inside a large body).

Our model does
not require to postulate a probability rule for the random localization
field, without any other justification than recovering the Born rule: the
correlations between the motions of the Bohmian position, guided as usually
by the wave function in the configuration space, are sufficient to ensure a
spatial localization of large massive objects, while the constant relaxation
towards quantum equilibrium \cite{Valentini-2012} automatically leads to the Born rule.

Pearle and Squires have remarked that the rate of collapse of GRW and CSL theories should be proportional to the mass, indicating a possible relation between collapse and gravitation \cite{Pearle-Squires-1993}. To connect our model with these theories, one can modify it  by assuming,  for instance, that the Bohmian position of each particle is the source of localization for this  particle only. The collapse then loses its collective properties and the model becomes more similar to GRW and CSL, maybe even equivalent. If it were equivalent, this would mean that the constant randomness of GRW and CSL theories, contained in their \textquotedblleft probability rule\textquotedblright , can also be interpreted within a deterministic dynamics in terms of random initial values of the Bohmian positions.  We have not explored this question.

\subsection{EEQT theory}

The Event-Enhanced-Quantum-theory (EEQT) \cite{Blanchard-et-al-2000} proposes a similar method to describe individual quantum systems and to explain why, in a measurement process, \textquotedblleft potential properties of a quantum system become actual\textquotedblright . It also enhances the standard quantum description of a system by replacing the usual space of states by a family of spaces, labelled with an index $\alpha$, representing the pure state of a classical system $C$.  An \textquotedblleft event\textquotedblright is defined by a change of the value of $\alpha$. Operators are labelled by two indices $\alpha$ and $\alpha '$, and not necessarily self-adjoint, as the non-Hermitian localization term we have introduced. A back action of the classical system is also introduced. Under these conditions, $\alpha$ plays a role in EEQT theory that is similar to the role of Bohmian positions in our model. The main difference is that the evolution of $\alpha$ is not deterministic,  but given by a Markov process.

\section{Conclusion}
\label{conclusion}

We have introduced two basic postulates: the source of gravitation is the
Bohmian density of particles, not the quantum density; the gravitational
coupling constant includes a small imaginary component. With these two
assumptions, predictions that are compatible with presently known facts are
obtained, including the appearance of single results in experiments. The dynamics is such that the mathematical objects (wave function
and positions) constantly follow the physical observations closely; there is no need
to update the value of the wave function in order to include new
information.\ For instance, if a sequence of measurements is performed on the
same quantum object, its state vector automatically includes the information
obtained in the previous measurements; there is no need to add a state
vector reduction by hand, or to keep empty components of the state vector.
As discussed in \cite{ABD} in more detail, the model remains compatible with a whole range of
possible interpretations and ontologies.

In this model, the quantum collapse is nothing but a consequence of the internal
cohesion of macroscopic objects and of their gravitational self-attraction
\cite{ABD}.\ The mutual attraction between the particles of the object
forces all Bohmian positions to remain grouped together, because they have
to occupy regions of the configuration space where the many particle wave
function does not vanish, a consequence of standard dBB\ theory.\ We then
assume that these positions collapse the state vector around them: in
equation (\ref{grav-11}), the source of gravitational attraction is the
Bohmian density, instead of the quantum density $D_{\Phi }(\mathbf{r})$
appearing in the Schr\"{o}dinger-Newton equation, discussed in detail for
instance in \cite{Bahrami-Bassi-et-al}. As early as in 1965, Bohm and Bub
\cite{Bohm-Bub} proposed to introduce a collapse dynamics involving hidden
variables (the components of a vector in the dual space of the Hilbert space
in their case). As mentioned in the introduction, Penrose \cite{Penrose-1996} suggested in 1996 that, when  a QSMDS\ involving
different spatial distribution of masses (and therefore different space-time
configurations) creates an energy fluctuation $\Delta E$, the QSMDS\ \
spontaneously decays in a time of the order of $\hbar /\Delta E$. The
spontaneous collapse arises because of an energy mismatch between two (or
more) components of the QSMDS.\ In this model, the primary origin of the collapse is
a mismatch between two densities of space, the quantum density and
the Bohmian density; this in turn creates a mismatch of gravitational energy in different components of the QSLMDS and achieves Penrose's scheme, but without any particular general relativistic effect.\ Recently, Tilloy has proposed a modification of
the GRW theory where the sources of a classical gravitational field are the
collapse space-time events of that theory \cite{Tilloy-2018}).

Depending on one's point of view, the role of the Bohmian positions can be seen as more, or less important, than in
standard dBB theory. In the dynamics, they certainly play a more active role than in dBB theory, where the positions do not appear in the dynamical equation giving the evolution of the
state vector, but just follow the spatial variations of the wave
function.
Here, the Bohmian positions act as mathematical attractors of the state vector  $\left\vert \Phi(t) \right\rangle$
through the gravitational term (including its small dissipative component in
$\varepsilon$). This introduces a nonlinearity in the
dynamics of $\left\vert \Phi(t) \right\rangle$. Nevertheless, as we have seen, in
most situations this change has very little effect on the evolution of $\left\vert \Phi(t) \right\rangle$ -- except in situations where QSMDS appear, which are then rapidly
projected by this term. The model therefore illustrates how the addition of a single additional variable to the standard equations, namely a point position in the configuration space, allows one to significantly enrich the dynamics and to take into account collapse situations.

From a purely interpretative point of view, one can see this continuous attraction as a pure mathematical ingredient to replace the stochastic fields of GRW and CSL theories, as well as their probability rule. One can then hold a view where the Bohmian positions are just mathematical objects creating this attraction, and where physical reality is directly represented, for instance, by the quantum density $D_{\Phi}(\mathbf r)$. But it is also perfectly possible to consider that all the individual Bohmian positions of the particles provide a direct representation of reality, as usual in dBB theory.

This model is in the line of calculations where gravity is treated
classically, within general relativity.\ This remains compatible with a
classical structure of space-time, in which the various quantum fields
(electromagnetic for instance) propagate (semi-classical gravity \cite%
{Struyve}); such schemes are sometimes useful in quantum cosmogenesis \cite%
{Peter-Pinho-et-al,Pinho-Pinto}. The circularity of the defintion of time in quantum theory \cite{Singh-2015} is avoided. A standard approach to semi-classical
gravity is to use a quantum average of the energy momentum tensor operator
to construct the Einstein tensor \cite{Moller,Rosenfeld,Rosenfeld-2}.\
Nevertheless, paradoxes may then arise: for instance, if a body is in a
quantum superposition of two locations, each localization of the body
attracts the other.\ Also, as discussed by Eppely and Hannah \cite%
{Eppley-Hannah}, one could in principle measure directly the modulus of the
wave function, and therefore obtain superluminal signaling.\ Other arguments
have been built, involving thought interference experiment, to discuss
possible inconsistencies, or to plead in favor of a semi-classical theory
of gravitation \cite{Baym-Ozawa,Bose-et-al,Belenchia-et-al,
Marletto-Vedral,Tilloy-2019}.\ In our model, as in that of Ref. \cite%
{Tilloy-2018}, the paradoxes arising from of delocalized sources of gravity
disappear: in each realization of an experiment, the source of gravitation
always remains localized in space (since it originates from the Bohmian
positions). Of course, in most situations (when no QSMDS\ occurs) it is
practically equivalent to take
the Bohmian positions, or the average quantum density of particles, as the source of gravity, due to the quantum equilibrium conditions.\ In this
sense, the predictions of this model are very similar to those of the theory
of semiclassical gravity proposed by Tilloy and Diosi \cite{Tilloy-Diosi},
the major difference being that their approach is based on a stochastic
spontaneous localization, while no random perturbation is invoked in the
present article.

At this stage, the model remains very elementary, in particular because its treatment of gravity remains simply
Newtonian, not Einsteinian: for instance, it does not include gravitational
waves. The hope is that the model could be an approximation of some more elaborate theory, compatible with
general relativity. One could also speculate about a
possible generalization to a quantum treatment of a gravitational field,
still having its sources in the Bohmian positions of the particles. One hope
could be to find a justification of the complex value of the coupling
constant by analogy with electromagnetic spontaneous emission, also taking
into account the intrinsic nonlinear character of general relativity. This,
of course, remains completely speculative. As it is, the model is definitely
in the line of a semi-classical treatment of gravity.

\textbf{Acknowledgments}: the author is grateful to Antoine Tilloy, Lajos Diosi, Philip Pearle and Nicolas Gisin for useful  comments and suggestions.

\end{document}